\documentclass[aps,prc,twocolumn,showpacs,amsmath,amssymb,nofootinbib]{revtex4-2}
\usepackage{graphicx}  
\usepackage{color}
\usepackage{times}
\usepackage{inputenc}
\usepackage{float}
\usepackage{bm}
\usepackage{ulem}
\usepackage{multirow}
\usepackage{float}
\usepackage{url}
\usepackage{natbib}
\usepackage{mathrsfs}
\usepackage{physics}
\usepackage{comment}
\usepackage[table,xcdraw]{xcolor}
\usepackage{booktabs,makecell}
\usepackage{comment}
\usepackage{amsmath}
\usepackage{pifont}

\usepackage[colorlinks=true,citecolor=blue,urlcolor=blue,linkcolor=blue]{hyperref}
\begin{document}
\title{Dark Matter Effects on the Curvature of Neutron Stars within the new Quarkyonic Model Coupled with Relativistic Mean Field Theory}
\author{Jeet Amrit Pattnaik$^1$}
\email{jeetamritboudh@gmail.com }
\author{D. Dey$^{2,3}$}
\author{R. N. Panda$^{1}$}
\author{M. Bhuyan$^{2}$}
\author{S. K. Patra$^{1}$}
\email{debabrat.d@iopb.res.in }

\affiliation{\it $^{1}$Department of Physics, Siksha $'O'$ Anusandhan, Deemed to be University, Bhubaneswar -751030, India}
\affiliation{\it $^{2}$Institute of Physics, Sachivalaya Marg, Bhubaneswar-751005, Odisha, India}
\affiliation{\it $^{3}$Homi Bhabha National Institute, Training School Complex, 
Anushakti Nagar, Mumbai 400094, India}
\date{\today}
\begin{abstract}
For the first time, we have analyzed the impact of dark matter (DM) on the curvature properties of quarkyonic neutron stars (NS) using a hybrid model based on quarkyonic-effective field theory within the relativistic mean-field (E-RMF) framework. This study focuses on the radial variation of key curvature components, including the Ricci scalar ($\cal{R}$), Ricci tensor ($\cal{J}$), Kretschmann scalar ($\cal{K}$), and Weyl tensor ($\cal{W}$), under different admixtures of dark matter. These curvature components provide critical insights into the spacetime geometry and gravitational field strength within the star. The model was applied to quarkyonic NS configurations, spanning both the canonical mass (1.4 $M_{\odot}$) and the maximum mass limit of the star. Two primary parameters were varied to capture the effects of matter transitions and quark confinement such as the transition density ($n_t$) and the QCD confinement scale ($\Lambda_{\rm cs}$), which governs the energy scale of quark confinement. These parameters, along with the inclusion of dark matter, allowed us to investigate the interplay between baryonic, quark, and DM components in shaping the curvature profiles. Our results reveal that the presence of quarkyonic matter (QM) and DM significantly alters the curvature within the star's interior. Central curvature values, particularly for the Kretschmann scalar ($\cal{K}$), Ricci scalar ($\cal{R}$) and Ricci tensor ($\cal{J}$), increase due to the higher central densities and pressures driven by DM and decreases in case of QM interactions. A stiffer EOS results in a smoother radial curvature profile, while a softer EOS, influenced by DM, redistributes curvature more dynamically. Additionally, the study explores the impact of DM and transition density on the compactness parameter is a critical measure of the gravitational field strength of the star. DM softens the EOS by reducing central pressure, leading to slightly less compact stars, while higher transition densities result in greater central pressures and compactness. These findings highlight the profound role of DM in shaping the internal and external curvature properties of quarkyonic neutron stars, offering new perspectives on the physics of compact objects in the presence of exotic matter.\\
\end{abstract}
\maketitle
\section{Introduction}
\label{intro}
Neutron stars (NSs) are one of the densest objects in the Universe, distinguished by their incredibly dense cores and robust crusts. These unique properties allow them to sustain persistent deformations, which can serve as potential sources of gravitational waves (GWs). Their extreme physical conditions make neutron stars invaluable for testing theories in high-density and strong-gravity regimes \cite{Shapiro_1983,Lattimer_2007}. However, our understanding of neutron star properties remains incomplete due to the limited knowledge of the equation of state (EOS) at densities significantly exceeding nuclear saturation. NSs predominantly consist of neutrons, along with smaller fractions of protons and leptons, but it is possible for exotic matter at high densities, such as hyperons, meson condensates, or even deconfined quarks, to exist within their cores.

As explained in our recent work \cite{dey2024}, recent observational advancements have greatly enhanced our understanding of NS properties. Precise measurements of NS masses exceeding $2 \, M_\odot$ and constraints on radii from gravitational wave events, like GW170817 \cite{PhysRevLett.119.161101, PhysRevLett.121.091102, PhysRevLett.121.161101, Capano2020}, as well as NICER’s X-ray observations of pulsars \cite{Riley_2019, Miller_2019}, have provided stringent limits on the equation of state (EOS) of dense matter. For example, the radius of a canonical $1.4 \, M_\odot$ NS is now constrained to be less than 13.5 km \cite{doi:10.1146/annurev-nucl-102711-095018,2016ARA&A..54..401O}. Similarly, to date, the heaviest black widow pulsars, such as PSR J0952-0607 and J0740+6620, provide additional constraints on NS mass, with values of $M = 2.35 \pm 0.17 M_\odot$ and $2.08 \pm 0.07 M_\odot$, respectively \cite{Romani_2022, Cromartie_2020}. These discoveries have motivated the development of new theoretical models, including the quarkyonic matter framework. This model suggests a transition between nucleonic and quark phases in the core of NSs, where quarks emerge as quasi-particles due to increasing densities, influencing the EOS and pressure behaviour \cite{PhysRevD.102.023021,dey2024}. Our previous study \cite{dey2024} adopts such an approach, examining the impact of DM and quark matter on the EOS and the resulting macroscopic characteristics of NSs. By varying model parameters, we tried to shed light on the interplay between nucleonic, quark, and DM components in shaping NS properties \cite{dey2024}.

The motivation behind the incorporation of dark matter (DM) into certain models is conceptualized primarily from the GW190814 event. It involved a compact binary coalescence where a black hole with a mass of 22.2 to 24.3 $M_\odot$ was paired with a secondary compact object weighing between 2.50 and 2.67 $M_\odot$, as reported with 90\% credible levels \cite{Abbott_2020}. This secondary object has sparked interest, as it could either be the lightest black hole ever observed or the most massive neutron star. Hence, it can be assumed that a certain amount of dark matter can be present inside the NS. Besides the extent of this DM capture determines its impact on NS characteristics such as their mass (M), radius (R), and tidal deformability ($\Lambda$) \cite{Sandin_2009,De_Lavallaz_2010,Kouvaris_2010,Ciarcelluti_2011,Leung_2011,AngLi_2012,DM1,Ellis_2018,Bhat_2019,Ivanytskyi_2019,DM3,DM2,mnras,das21}. In this work, we have considered the non-annihilating WIMPs as the DM candidate, noting that their inclusion softens the EOS, leading to a reduction in the NS’s M and R. To compute the equation of state, we have chosen the Effective-Relativistic mean-field (E-RMF) theory, a widely used framework, which models the interactions between nucleons and mesons, providing a consistent and effective approach for describing dense nuclear matter. By extending the E-RMF formalism to include quark matter and DM, it becomes possible to analyze the combined effects of these components on NS properties. In this work, we conduct a study on the curvature analysis of quarkyonic NSs influenced by DM, with the G3 \cite{G3} and IOPB-I \cite{IOPB-I} parameter sets. These parameters have the merit of reproducing most of the known properties of finite nuclei, such as binding energy, root mean square charge radius etc, not only for beta-stable nuclei but also for nuclei that are away from the stability line. Also, the prediction of the mass and radius of neutron star comes in closure within the recent observational limits along with the tidal deformability in a binary neutron star merger \cite{G3, IOPB-I, dey2024, patt1, patt3}. Our study aims to provide deeper insights into the interplay between DM and spacetime curvature in these compact objects.

The spacetime geometry near neutron stars plays a crucial role in constraining the equation of state (EOS) of dense matter. Scalar and tensor quantities, such as the Ricci scalar, Kretschmann scalar, and the Ricci and Weyl curvature tensors, characterize the curvature of spacetime within and around neutron stars. These quantities describe how spacetime is warped by the star’s immense mass and compactness. By analyzing the effects of curvature on astrophysical phenomena—such as binary neutron star mergers, light bending, the redshift of electromagnetic radiation emitted from the neutron star’s surface, and time dilation—one can derive constraints on the EOS. As noted in Refs. \cite{Psaltis_2008, Xiao_2015}, spacetime curvature can be quantified using the compactness parameter, which ranges from zero to one, representing the transition from flat spacetime to the extreme curvature at a black hole's event horizon. Direct measurements of spacetime curvature are achievable using techniques like atom interferometry \cite{Rosi_2015, das21}, offering a promising approach to further refine our understanding of the EOS.

From a mathematical and physical point of view, the Riemann tensor measures the curvature of space and has 20 independent components in four dimensions. From it, one can derive other important quantities such as the Kretschmann scalar (a measure of curvature magnitude), the Ricci tensor (a simplified version focusing on volumetric changes), and the Ricci scalar (a single value summarizing the Ricci tensor). The Weyl tensor is another derivative, capturing shape distortions by removing all volumetric effects. Together, these tensors describe how tidal forces affect a body's shape and volume, with the Riemann tensor combining both aspects. More details can be found in Refs. \cite{carroll_2019,das21}. Further, it can be seen in the curvature analysis done by Eksi et al. \cite{Kazim_2014} to understand the gravity in the GR framework. There he found the GR is less rigorously constrained across the star than the EOS. Again, Xiao et al. \cite{Xiao_2015} have confirmed the strong dependency of symmetry energy on curvature in low-mass NSs, which significantly fade away moving towards the massive ones. In addition to that, the works of Das et al. \cite{das21}, ensure that a small amount of dark matter may present inside the NS, however, more percentages of DM will make the star unstable. On a different note, in our recent study \cite{das23}, we have also analysed the effect of anisotropy on the surface curvature of NS and given various approximate relations for surface curvature-tidal deformability (SC - $\Lambda$), surface curvature-moment of inertia (SC - $\bar I$) for the canonical star.

This paper is organized as follows: Section \ref{formalism} presents the theoretical framework, detailing the E-RMF approach and its extensions to incorporate quark matter and dark matter. Section \ref{results} covers the analysis and findings of the study, followed by the concluding remarks in Section \ref{Conclusions}.
\section{Theoretical framework}
\label{formalism}
\subsection{Nuclear model}
\label{a}
The effective relativistic mean-field (E-RMF) formalism has proven to be a reliable framework for understanding both finite nuclei and infinite nuclear matter (NM) \cite{patt1, patt2, patt3, patt4}. The parameters of the E-RMF model are calibrated using a wide range of experimental and empirical data. Over 200 parameter sets have been developed in the literature to replicate various experimental and observational findings \cite{PhysRevC.55.540,PhysRevC.63.044303,PhysRevC.74.045806,PhysRevC.70.058801,LALAZISSIS200936,PhysRevC.82.055803,PhysRevC.82.025203,PhysRevC.84.054309,PhysRevC.85.024302,IOPB-I,PhysRevC.102.065805}. For this study, we employed the E-RMF model, which has been detailed extensively in Refs. \cite{das21, E-RMF2, E-RMF3, E-RMF4, E-RMF5}. The inclusion of leptonic contributions is also critical to ensure the stability of neutron stars. Accordingly, the energy density ($\mathcal{E}_{\rm BM}$) and pressure ($P_{\rm BM}$) for a system comprising NM and leptons are determined using stress-energy tensor techniques, as described in Ref. \cite{E-RMF5}.
\begin{eqnarray}
\label{eq:eden}
{\cal E}_{\rm BM} & = & \sum_{i=p,n} \frac{g_s}{(2\pi)^{3}}\int_{0}^{k_{f_{i}}} d^{3}k\, \sqrt{k^{2} + M_{\rm nucl.}^{*2}}\nonumber\\
&&
+n_{b} g_\omega\,\omega+m_{\sigma}^2{\sigma}^2\Bigg(\frac{1}{2}+\frac{\kappa_{3}}{3!}\frac{g_\sigma\sigma}{M_{\rm nucl.}}+\frac{\kappa_4}{4!}\frac{g_\sigma^2\sigma^2}{M_{\rm nucl.}^2}\Bigg)
\nonumber\\
&&
 -\frac{1}{4!}\zeta_{0}\,{g_{\omega}^2}\,\omega^4
 -\frac{1}{2}m_{\omega}^2\,\omega^2\Bigg(1+\eta_{1}\frac{g_\sigma\sigma}{M_{\rm nucl.}}+\frac{\eta_{2}}{2}\frac{g_\sigma^2\sigma^2}{M_{\rm nucl.}^2}\Bigg)
 \nonumber\\
&&
 + \frac{1}{2} (n_{n} - n_{p}) \,g_\rho\,\rho
 -\frac{1}{2}\Bigg(1+\frac{\eta_{\rho}g_\sigma\sigma}{M_{\rm nucl.}}\Bigg)m_{\rho}^2
 \nonumber\\
 && 
-\Lambda_{\omega}\, g_\rho^2\, g_\omega^2\, \rho^2\, \omega^2
+\frac{1}{2}m_{\delta}^2\, \delta^{2}
\nonumber\\
 && 
+\sum_{j=e,\mu}  \frac{g_s}{(2\pi)^{3}}\int_{0}^{k_{F_{j}}} \sqrt{k^2 + m^2_{j}} \, d^{3}k,
\end{eqnarray}
and
\begin{eqnarray}
\label{eq:press}
P_{\rm BM} & = & \sum_{i=p,n} \frac{g_s}{3 (2\pi)^{3}}\int_{0}^{k_{f_{i}}} d^{3}k\, \frac{k^2}{\sqrt{k^{2} + M_{\rm nucl.}^{*2}}} \nonumber\\
&& - m_{\sigma}^2{\sigma}^2\Bigg(\frac{1}{2} + \frac{\kappa_{3}}{3!}\frac{g_\sigma\sigma}{M_{\rm nucl.}} + \frac{\kappa_4}{4!}\frac{g_\sigma^2\sigma^2}{M_{\rm nucl.}^2}\Bigg)+ \frac{1}{4!}\zeta_{0}\,{g_{\omega}^2}\,\omega^4 
\nonumber\\
&&
+\frac{1}{2}m_{\omega}^2\omega^2\Bigg(1+\eta_{1}\frac{g_\sigma\sigma}{M_{\rm nucl.}}+\frac{\eta_{2}}{2}\frac{g_\sigma^2\sigma^2}{M_{\rm nucl.}^2}\Bigg)
\nonumber\\
&&
+ \frac{1}{2}\Bigg(1+\frac{\eta_{\rho}g_\sigma\sigma}{M_{\rm nucl.}}\Bigg)m_{\rho}^2\,\rho^{2}-\frac{1}{2}m_{\delta}^2\, \delta^{2}+\Lambda_{\omega} g_\rho^2 g_\omega^2 \rho^2 \omega^2
\nonumber\\
&&
+\sum_{j=e,\mu}  \frac{g_s}{3(2\pi)^{3}}\int_{0}^{k_{F_{j}}} \frac{k^2}{\sqrt{k^2 + m^2_{j}}} \, d^{3}k,
\end{eqnarray}
where $g_s$ and $M$ represent the spin degeneracy and mass of the nucleon. The $m_\sigma$, $m_\omega$, $m_\rho$, and $m_\delta$ are the masses and $g_\sigma$, $g_\omega$, $g_\rho$, and $g_\delta$ are the coupling constants for the $\sigma$, $\omega$, $\rho$, and $\delta$ mesons respectively. Other couplings such as $\kappa_3$, $\kappa_4$, $\zeta_0$ are for the self-interactions, and $\eta_1$, $\eta_2$, $\eta_\rho$, and $\Lambda_\omega$ are the cross-couplings between mesons \cite{ERMF6, Serot1992, ERMF7, E-RMF8, E-RMF9, E-RMF5, IOPB-I}.
\subsection{Quarkyonic Model}
\label{b}
We now delve into the quarkyonic model introduced by McLerran and Reddy \cite{PhysRevLett.122.122701}. This phenomenological framework proposes that, within the core of neutron stars (NS) where the density far exceeds nuclear saturation density, nucleons disintegrate into quarks. The transition from NM to the quarkyonic phase is characterized by a sharp increase in pressure, driven by the population of quarks in low-momentum states as the baryon density surpasses a critical threshold, known as the transition density. In this regime, low-momentum states are treated as quarks, while high-momentum states near the Fermi surface remain nucleonic. The Fermi surface momentum scales are of the order of the QCD confinement scale, $\Lambda_{\rm cs}$, resulting in the formation of nucleons as bound states of quarks.

The schematic model by McLerran and Reddy \cite{PhysRevLett.122.122701} was later extended and refined by Zhao and Lattimer \cite{PhysRevD.102.023021}, who incorporated beta-equilibrium and charge neutrality for quarkyonic matter. In their modification, nucleons interact through a density-dependent potential energy, calibrated to match specific properties of uniform NM. Moreover, they introduced chemical equilibrium between nucleons and quarks to define the relationship between the nucleonic ($k_{f_{n,p}}$) and quark ($k_{f_{u,d}}$) Fermi momenta, distinguishing this refined quarkyonic model. In this framework, nucleons occupy a finite Fermi shell, with a minimum Fermi momentum ($k_{f0_{(n,p)}}$) and an upper Fermi momentum ($k_{f_{n,p}}$). Similarly, the Fermi momenta for $u$ and $d$ quarks are denoted as $k_{f_u}$ and $k_{f_d}$, respectively. A comprehensive analysis of this model and its implications can be found in our recent work \cite{dey2024}.

The quarks energy density  ${\cal E}_{\rm QM}$ and pressure $P_{\rm QM}$ can be expressed as \cite{PhysRevD.102.023021,dey2024, dey2024},
\begin{eqnarray}
{\cal E}_{\rm QM}&=& \sum_{j=u,d}\frac{g_s N_c}{(2\pi)^3}\int_0^{k_{f_{j}}}k^2\sqrt{k^2 + m_{j}^2 }\, d^3k,
\end{eqnarray}
\begin{eqnarray}
P_{\rm QM} &=& \mu_{u} n_{u} + \mu_{d} n_{d} - \epsilon_{QM} \, .
\end{eqnarray} 
\subsection{Dark Matter Model}
\label{c}
The dark matter Lagrangian density \cite{DM1,DM2,DM3,dey2024} is constructed based on the interaction DM particles with both nucleons and quarks channelling through Higgs exchange, which is defined as,
\begin{eqnarray}
{\cal{L}}_{\rm DM} &=& \bar{\chi} \left[ i \gamma^\mu \partial_\mu - M_\chi + y h \right] \chi 
+ \frac{1}{2} \partial_\mu h \partial^\mu h \nonumber \\
&& - \frac{1}{2} M_h^2 h^2 + f \frac{M_{\text{nucl.}/u/d}}{v} \bar{\psi} h \psi ,
\label{eq:LDM}
\end{eqnarray}

In this model, $\chi$ and $\psi$ represent the wave functions of the Dark Matter (DM) particle and nucleons, respectively. The interaction between the Higgs boson and nucleons follows a Yukawa-type coupling, characterized by the coupling constant $f$, which corresponds to the proton-Higgs form factor. We assume the Neutralino as the DM particle with a mass $M_\chi$ of 200 GeV. The parameters $y$ and $f$ are chosen as 0.07 and 0.35, respectively, based on constraints derived from experimental and empirical data \cite{DM5}. Additionally, the mass of the Higgs boson ($M_h$) is fixed at 125 GeV, while its vacuum expectation value ($v$) is set to 246 GeV.

Using the mean-field approximation the energy density and pressure for the DM can be expressed as  \cite{DM1, DM3, DM2, mnras, DM11,DM12, DM13}
\begin{eqnarray}
{\cal{E}}_{\rm DM}& = & \frac{2}{(2\pi)^{3}}\int_0^{k_f^{\rm DM}} d^{3}k \sqrt{k^2 + (M_\chi^\star)^2} + \frac{1}{2}M_h^2 h_0^2 ,
\label{eq:EDM}
\end{eqnarray}
\begin{eqnarray}
P_{\rm DM}& = &\frac{2}{3(2\pi)^{3}}\int_0^{k_f^{\rm DM}} \frac{d^{3}k k^2} {\sqrt{k^2 + (M_\chi^\star)^2}} - \frac{1}{2}M_h^2 h_0^2 ,
\label{eq:PDM}
\end{eqnarray}
where $k_f^{\rm DM}$ is the DM Fermi momentum.\\

\subsection{The Complete Model for Dark Matter-admixed Quarkyonic Neutron Star}
\label{d}
The total energy density and pressure of a DM-admixed quarkyonic star are then given by:

\begin{eqnarray}
\cal{E} &=& {\cal{E}}_{\rm BM} + {\cal{E}}_{\rm QM} + {\cal{E}}_{\rm DM},
\label{eq:effm_total}
\end{eqnarray}

\begin{eqnarray}
P &=& P_{\rm BM} + P_{\rm QM} + P_{\rm DM}.
\label{eq:press_total}
\end{eqnarray}
\noindent
where ${\cal{E}}_{\rm BM}$, ${\cal{E}}_{\rm QM}$, and ${\cal{E}}_{\rm DM}$ represent the energy densities of nucleonic matter, quark matter, and dark matter, respectively, and $P_{\rm BM}$, $P_{\rm QM}$, and $P_{\rm DM}$ are their corresponding pressures. These are obtained from the respective Lagrangians and combined to form total ${\cal{E}}$ and $P$, which are further used in the TOV equations to evaluate the NS properties. These sharp quark-nucleon phase transitions generally give discontinuity in EOS and are made continuous using Gibb's criteria. A full detailed procedure is present in our previous work \cite{dey2024, patra07}.

 \subsection{Mass and Radius of the NS}
   We determine neutron star (NS) observables such as \(M\) and \(R\) by solving the Tolman-Oppenheimer-Volkoff (TOV) equations. For this, the equations of state (EOSs) of NSs incorporating dark matter (DM) are utilized as inputs to the TOV equations \citep{TOV1,TOV2}, which are expressed as:
    \begin{eqnarray}
    \frac{dP_{tot.}(r)}{dr}= - \frac{(P_{tot.}(r)+{\cal{E}}_{tot.}(r))(m(r)+4\pi r^3 P_{tot.}(r))}{r(r-2m(r))}, \nonumber\\
    \label{TOV1}
    \end{eqnarray}
    and 
    \begin{eqnarray}
    \frac{dm(r)}{dr}=4\pi r^2 {\cal{E}}_{tot.}(r).
    \label{TOV2}
    \end{eqnarray}
   Here, \({\cal{E}}_{tot.}(r)\) and \(P_{tot.}(r)\) represent the total energy density and pressure, respectively, as functions of the radial distance \(r\). The term \(m(r)\) corresponds to the gravitational mass enclosed within a radius \(r\). These coupled equations are solved simultaneously to determine the mass and radius of the neutron star for a given central density.
\subsection{Mathematical expressions for various curvatures}
\label{curvature}
In the study of NS and general relativity in general four kinds of curvature help us to characterize the structure of space-time both inside as well as outside of stars. The four different measures of curvatures are Ricci scalar ($\cal{R}$), Ricci tensor ($\cal{J}$), Kretschmann scalar ($\cal{K}$) and Weyl tensor ($\cal{W}$). The mathematical expression can be given as, 
    the Ricci scalar,
    \begin{equation}
    {\cal R}(r)=8\pi\bigg[{\cal{E}}_{tot.}(r) -3 P_{tot.}(r)\bigg],
    \label{RS}
    \end{equation}
    the square root of the full contraction of the Ricci tensor is defined as 
    \begin{equation}
    {\cal J}(r) \equiv \sqrt{{\cal R}_{\mu \nu} {\cal R}^{\mu \nu}} = \bigg[(8\pi)^2 \left[ {\cal{E}}_{tot.}^2(r) + 3P_{tot.}^2(r)\right]\bigg]^{1/2},
    \label{RT}
    \end{equation}
    the Kretschmann scalar is defined as the square root of the full contraction of the Riemann tensor.
    \begin{eqnarray}
    {\cal{K}}(r)&\equiv&\sqrt{{\cal{R}}^{\mu\nu\rho\sigma}{\cal{R}}_{\mu\nu\rho\sigma}}
    \nonumber\\
    &&
    =\bigg[(8\pi)^2[3{\cal{E}}_{tot.}^2(r)+3P_{tot.}^2(r)
    +2P_{tot.}(r){\cal{E}}_{tot.}(r)]
    \nonumber\\
    &&
    -\frac{128{\cal{E}}_{tot.}(r)m(r)}{r^3}+\frac{48m^2(r)}{r^6}\bigg]^{1/2},
    \label{KS}
    \end{eqnarray}
    and the square root of the full contraction of the Weyl tensor
    \begin{equation}
    {\cal W}(r) \equiv \sqrt{{\cal C}^{\mu \nu \rho \sigma }{\cal C}_{\mu \nu \rho \sigma}} = \bigg[\frac43 \left( \frac{6m(r)}{r^3} - 8\pi {\cal{E}}_{tot.}(r) \right)^2\bigg]^{1/2}.
    \label{WT}
    \end{equation}
    Where  ${\cal{E}}_{tot.}$, $P_{tot.}$, $m(r)$ are the energy density, pressure, and mass of the NS as a function of radius respectively. The $\cal{K}$ and $\cal{W}$ curvature are present inside and outside of the star i.e. in a vacuum. Contrary to this $\cal{R}$ and $\cal{J}$ are only confined within the star. More details can be found in Refs. \cite{Kazim_2014, Xiao_2015,das21}.
    
\section{Results and Discussions}
\label{results}
This section highlights the influence of various exotic constituents within a neutron star, analyzed through different types of spacetime curvatures. In our previous work \cite{dey2024}, we introduced the quarkyonic framework characterized by two free parameters: the transition density ($n_t$) and the QCD confinement scale ($\Lambda_{\rm cs}$). Here, $n_t$ governs the onset of quark formation, while $\Lambda_{\rm cs}$ serves as a cutoff between nucleons and quarks in the momentum scale. Additionally, the effects of dark matter (DM) are incorporated into the quarkyonic star model by introducing different values of the DM Fermi momentum ($k_f^{\rm DM}$), allowing for a comprehensive analysis of its impact.

\begin{figure}
\centering
\includegraphics[width=1\columnwidth]{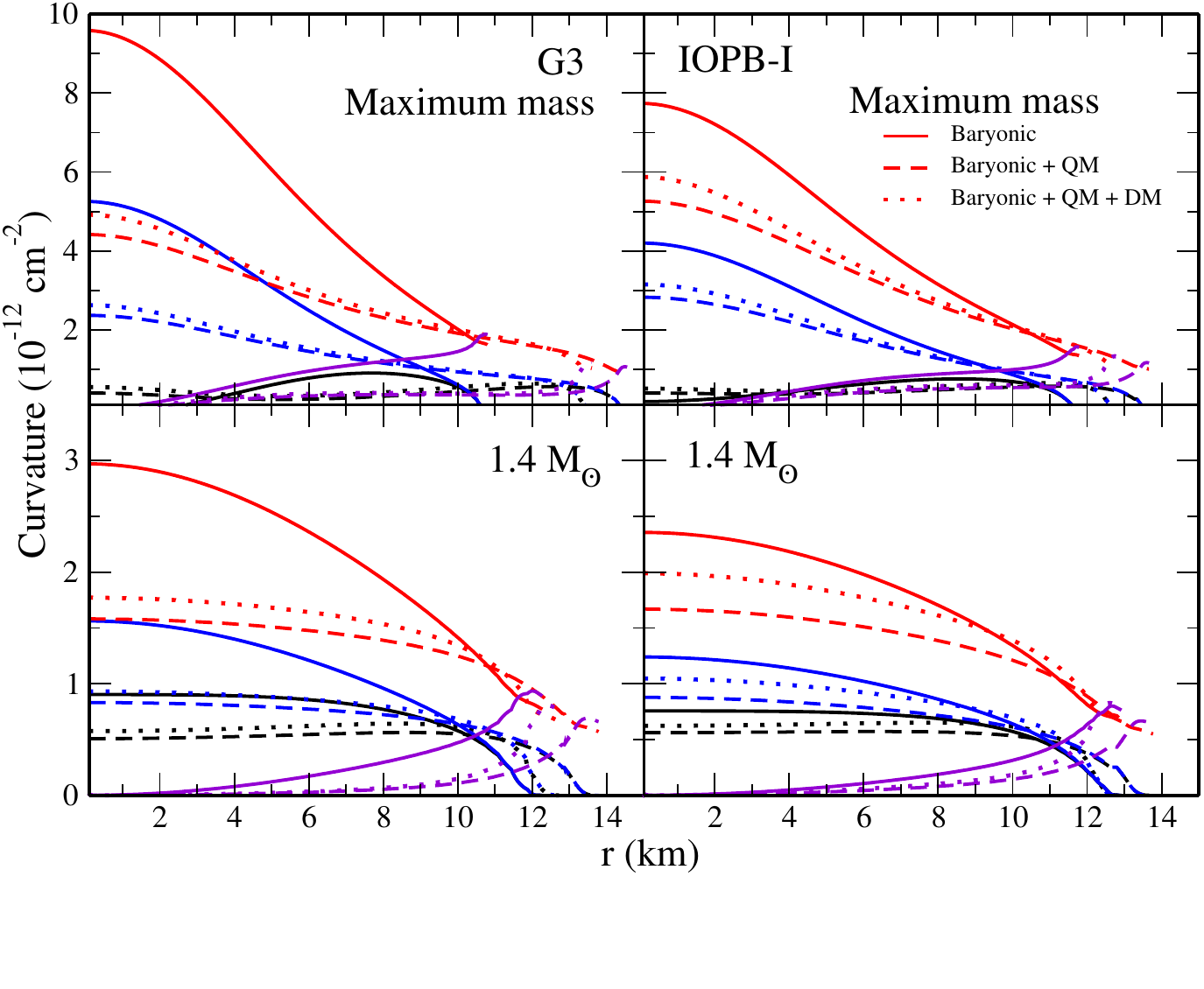}
\caption{Different types of curvatures such as  $\cal{R}$ (black), $\cal{J}$ (blue), $\cal{K}$ (red), and $\cal{W}$ (violet) with radius (r (km)) for (a) baryonic (b) quarkyonic ($\rm n_t$ = 0.3 $\rm fm^{-3}$, $\Lambda_{\rm cs}$ = 800 $\rm MeV$, $\rm k_f^{DM}$ = 0.00 $\rm GeV$) and (c)  DM admixed with quarkyonic ($\rm n_t$ = 0.3 $\rm fm^{-3}$, $\Lambda_{\rm cs}$ = 800 $\rm MeV$, $\rm k_f^{DM}$ = 0.03 $\rm GeV$) for NM parameter G3 and IOPB-I. } 
\label{fig1}
\end{figure}
\begin{figure}
\centering
\includegraphics[width=1\columnwidth]{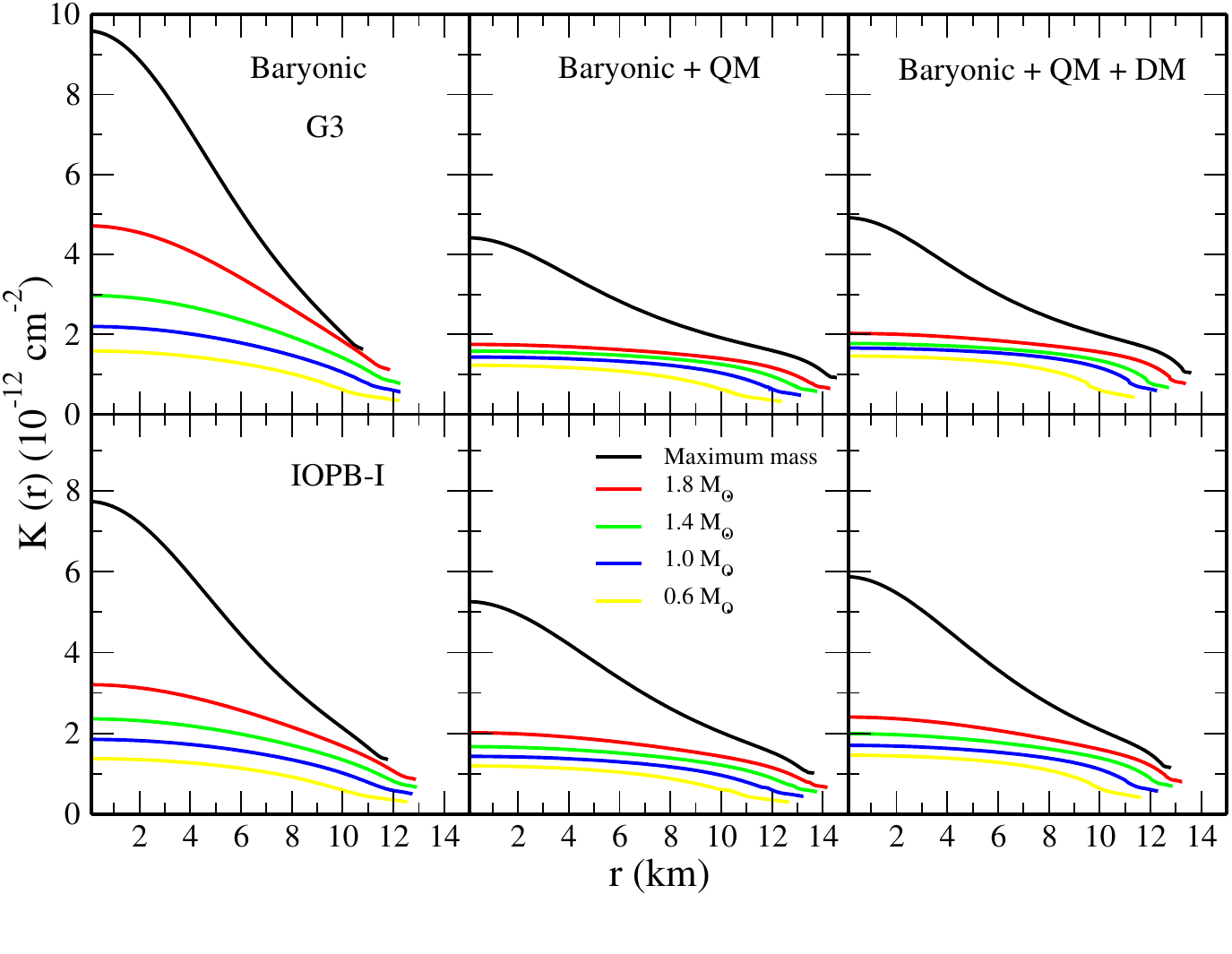}
\caption{The radial variation of ${\cal{K}}(r)$ curvature for (a) baryonic (b) quarkyonic ($\rm n_t$ = 0.3 $\rm fm^{-3}$, $\Lambda_{\rm cs}$ = 800 $\rm MeV$, $\rm k_f^{DM}$ = 0.00 $\rm GeV$) and (c)  DM admixed with quarkyonic ($\rm n_t$ = 0.3 $\rm fm^{-3}$, $\Lambda_{\rm cs}$ = 800 $\rm MeV$, $\rm k_f^{DM}$ = 0.03 $\rm GeV$) for NM parameter G3 and IOPB-I. Here, we have presented a mass range from 0.6 (M$_\odot$) to maximum mass (M$_\odot$).
}\label{fig2}
\end{figure}

In this analysis, we have computed radial variation of various curvature quantities such as  $\cal{R}$, $\cal{J}$, $\cal{K}$, and $\cal{W}$  for three compact objects, namely, baryonic star, quarkyonic star and dark matter admixed quarkyonic star, which is depicted in Fig. \ref{fig1}. All the curvatures have the maximum value at the core of the star for each scenario. However, the  Weyl tensor reached negative values. We noticed a larger gap in all curvatures for quarkyonic stars than baryonic, ones due to their stiffer equation of state and high compactness. After adding the dark matter, the curvature values little increases, but the effect is eventually ruled out while moving towards the surface. Another notable difference is the Ricci scalar, which is negative for maximum mass while it is positive for canonical star. This negative behaviour shows the dominance of pressure over the energy density in a star. Hence the exploration of the equation of state is much more needed. Further mathematically, $\cal{K}$ and $\cal{W}$ can be similar at the surface only when the ${\cal{E}}_{tot.} $ and $P_{tot.} $ will be zero, while the other two $\cal{J}$ and $\cal{R}$ approach to zero. Additionally, it is reported in the works of Das et al. \cite{das21} that the density becomes diffused while moving from the core to the surface, resulting in $\cal{W}$ reaching its maximum. A similar kind of trend is observed here. Further, we notice the radial variation of curvature is significant in quarkyonic and DM admixed quarkyonic case for maximum mass, while the change is very minimal, i.e. almost flat curve for the canonical star. Among these curvatures, we analyse Kretschmann scalar (\({\cal{K}}(r)\)) within the neutron star, quarkyonic star, as well as DM admixed quarkyonic star as illustrated in Fig. \ref{fig2} for various stellar masses ranging from 0.6 (M$_\odot$) to maximum mass (M$_\odot$). To examine the curvature's dependence on the radial coordinate, the transition density $\rm n_t$ = 0.3 $\rm fm^{-3}$, and QCD confinement parameter $\Lambda_{\rm cs}$ = 800 $\rm MeV$ are considered for a quarkyonic star. Besides to see the potential effects of DM, we have fixed the dark matter Fermi momentum $\rm k_f^{DM}$ at 0.03 GeV for a particular star. 

The Kretschmann scalar, ${\cal K} (r)$, exhibits notable radial variation near the core, rising sharply with increasing neutron star compactness. As the stellar mass increases, central curvature intensifies, reaching its peak at the maximum mass configuration. This pattern reflects a concentrated gravitational field and curvature within the core regions of baryonic neutron stars. Introducing quarkyonic matter leads to a stiffer equation of state (EOS) compared to the purely baryonic case, supporting higher masses and central pressures, which in turn result in lesser central curvature values. The smooth variation in the curve reflects the quark-nucleon crossover, where the transitions occurred gradually between nucleonic and quark matter phases, preventing abrupt changes in the pressure and density profiles. Additionally, the stiffer EOS leads to slightly larger radii for the same mass, redistributing the curvature more evenly across the stellar radius. Adding dark matter softens the quarkyonic EOS, reducing the central pressure and, further, the core curvature starts rising. The outer layers exhibit slightly higher curvature values compared to the purely quarkyonic case, indicating the influence of DM on redistributing mass-energy within the star. Similarly DM affects all the curvature parameters such as $\cal{R}$, $\cal{J}$, $\cal{K}$, and $\cal{W}$ of the NS considerably. Further, both G3 and IOPB-I parametrizations show consistent trends in the curvature profiles, though slight differences arise due to the inherent stiffness of the EOS models.

\begin{figure}
\centering
\includegraphics[width=1\columnwidth]{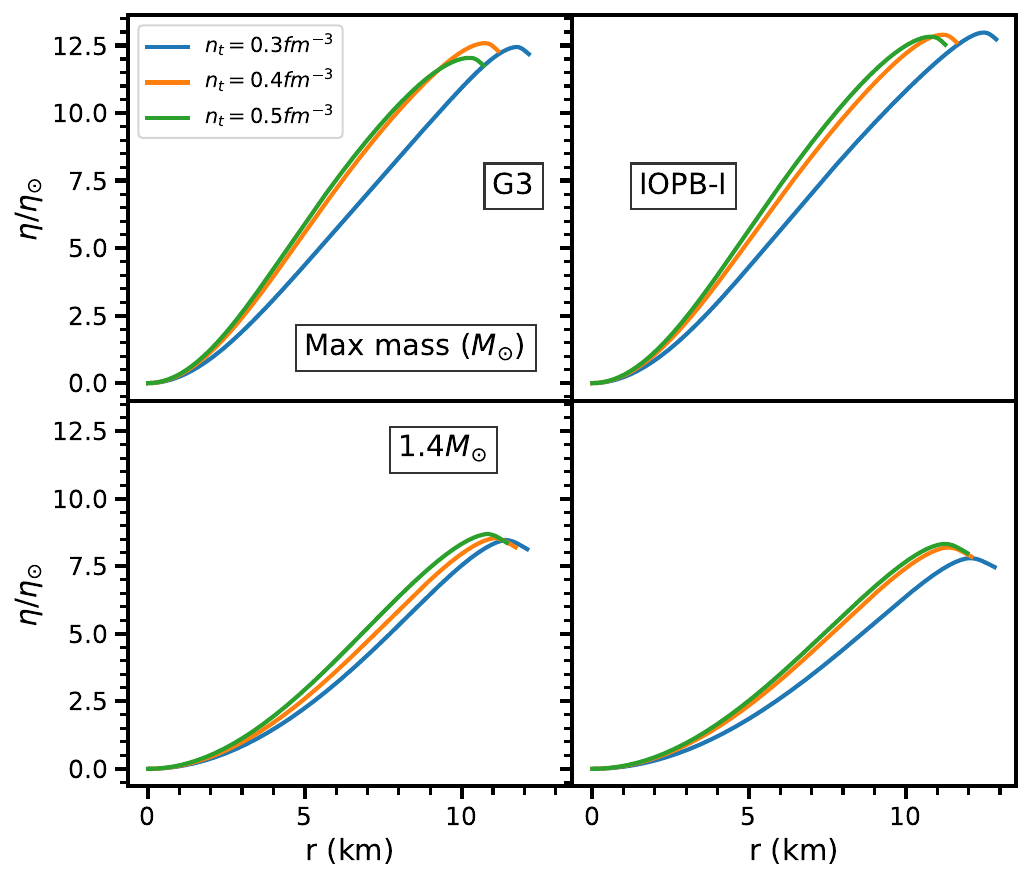}
\caption{ The ratio of the compactness of NS and the Sun with the variation of NS radius at different transition densities, such as $n_t = 0.3, 0.4, 0.5 \rm fm^{-3}$, keeping the DM Fermi momentum $\rm k_f^{DM}$ and confinement scale parameter ($\Lambda_{\rm cs}$) at 0.03 $\rm GeV$ and 800 $\rm MeV$ respectively for both maximum and canonical star with G3 and IOPB-I sets.}
\label{fig3}
\end{figure}

In Fig. \ref{fig3}, we have obtained the ratio of compactness between neutron stars (NS) and the Sun with the change in NS radius at various transition densities, \( n_t = 0.3, 0.4, 0.5 \, \mathrm{fm^{-3}} \), using the G3 and IOPB-I sets. The dark matter Fermi momentum $\rm k_f^{DM}$ at 0.03 GeV is held constant throughout all the cases. Generally, the compactness ($\eta=\frac{GM}{r c^2}$) quantifies the density of a star. Neutron stars (NSs), having significantly greater mass and smaller radii than the Sun, exhibit compactness values approximately \(10^5\) times higher than that of the Sun. In the present analysis, we noticed, that the compactness ratio is affected by the transition density, such as higher transition densities (\( n_t = 0.5 \, \mathrm{fm^{-3}} \)) resulted in higher compactness ratios, indicating a stiffer equation of state (EOS) and larger radii. Similarly, at (\( n_t = 0.3 \, \mathrm{fm^{-3}} \)), the ratio is lesser indicating softer EOS. Additionally, compactness is higher for the maximum-mass neutron star compared to the canonical star. Further, the compactness gradually increases while moving towards the outer layers. Moreover, one can conclude that the stars having a larger quarkyonic core, are the most compact objects in nature.

\begin{figure}
\centering
\includegraphics[width=1\columnwidth]{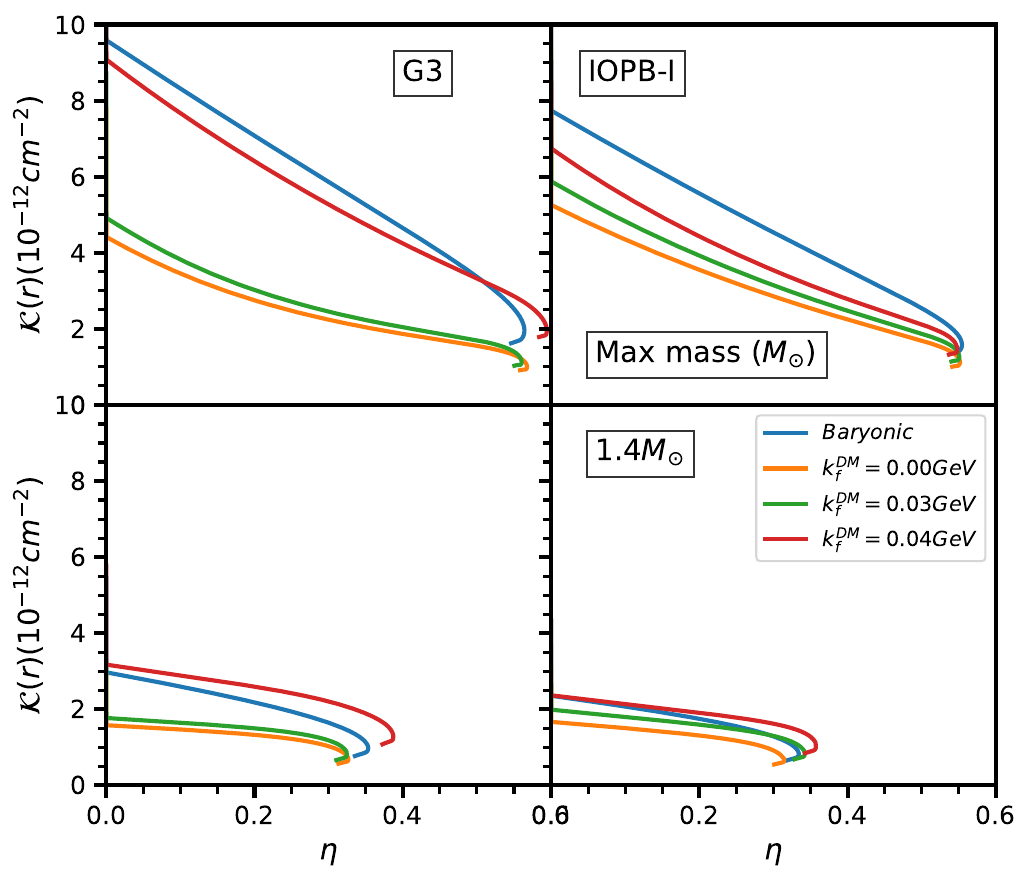}
\caption{The variation of Kretschmann scalar curvature (${\cal{K}}(r)$) of NS for the compactness of baryonic and DM admixed quarkyonic stars having DM Fermi momentum $\rm k_f^{DM}$ = 0.00, 0.03, and 0.04 $\rm GeV$ with DM for G3 and IOPB-I. Here the transition density ($\rm n_t$) and confinement scale parameter ($\Lambda_{\rm cs}$) are kept at 0.3 $\rm fm^{-3}$ and 800 $\rm MeV$ respectively, for both maximum and canonical star.}
\label{fig4}
\end{figure}

As compactness plays a crucial role in the curvature analysis, we study the variation of the Kretschmann scalar curvature (${\cal{K}}(r)$) inside neutron stars (NS) as a function of the compactness parameter, for maximum mass and canonical mass star for both baryonic matter and DM admixed quarkyonic matter scenarios which are illustrated in Fig. \ref{fig4}. The Kretschmann scalar curvature for a purely baryonic neutron star serves as the basis for understanding the impact of dark matter (DM) on the curvature. The addition of quarkyonic matter to the core stiffens the equation of state resulting in a drastic fall of curvature at the core of the star, which further increases with the inclusion of dark matter. This is due to softening the EOS and leading to increased central density. It can be seen from the figure that the higher values of the Kretschmann scalar (curvature) for fixed compactness are only when dark matter is present. As the Fermi momentum of dark matter increases, the curvature profile exhibits more pronounced deviations from the baryonic star, showing the gravitational impact of dark matter. Both G3 and IOPB-I, equation of states exhibit similar overall trends, with an increase in curvature as compactness rises.  Notably, the variation in \({\cal{K}}(r)\) remains relatively small up to the canonical mass compared to the changes observed at the maximum mass of the star. Also, we noticed, that for a canonical star, dark matter Fermi momentum $\rm k_f^{DM}$ at 0.04 GeV has a higher curvature value as compared to the baryonic case, with both the parameters.
\begin{figure}
\centering
\includegraphics[width=1\columnwidth]{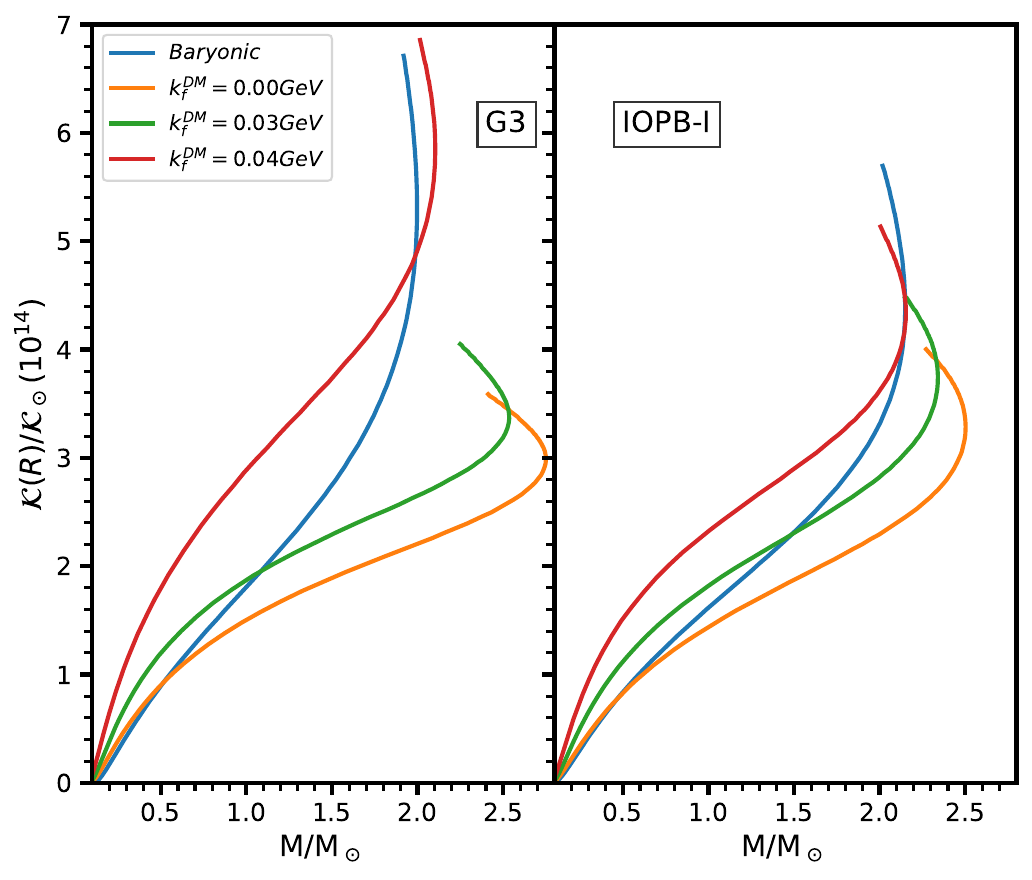}
\caption{The ratio of the Kretschmann scalar surface curvature of NS and the Sun \({\cal{K}}(R)/\cal{K}_{\odot}\) with the mass for baryonic and DM admixed quarkyonic stars having DM Fermi momentum $\rm k_f^{DM}$ = 0.00, 0.03, and 0.04 $\rm GeV$ for G3 and IOPB-I forces. Here the transition density ($\rm n_t$) and confinement scale parameter ($\Lambda_{\rm cs}$) are kept at 0.3 $\rm fm^{-3}$ and 800 $\rm MeV$ respectively.}
\label{fig5}
\end{figure}

We have presented the Kretschmann scalar curvature at the surface of NS in Fig. \ref{fig5}. Elaborately, the Figure illustrates the variation of \({\cal{K}}(R)/\cal{K}_{\odot}\) with the neutron star's mass, where the Sun's surface curvature (\(\cal{K}_{\odot}\)) is \(3.06 \times 10^{-27}~\text{cm}^{-2}\). From our analysis, it is observed that a quarkyonic star predicts lesser surface curvature $\cal{K} (R)$ than the pure baryonic case, due to its stiffer nature. However, the effect alters after adding the dark matter. The effect intensifies with the process of increasing DM percentages. In support of our work, we are getting a similar trend as obtained by Das et al. \cite{das21}, that the ratio ${\cal{K}}(R)$/$\cal{K}_{\odot}$, increases with the inclusion of dark matter (DM). Further, the G3 parameter set, which produces a softer equation of state (EOS) compared to IOPB-I, results in larger surface curvature. Greater curvature indicates stronger spacetime warping, directly influenced by the percentage of DM within the neutron star.

\section{Conclusions} \label{Conclusions}
In summary, we conducted a detailed investigation into the curvature properties of quarkyonic stars, emphasizing the influence of dark matter within the effective relativistic mean-field (E-RMF) framework. Our study explored the gravitational curvature parameters, such as the Kretschmann scalar ${\cal K}$ and Ricci scalar ${\cal R}$, which provide insight into the strong-field regime of general relativity and the internal structure of neutron stars. The calculations were carried out using two widely studied nuclear parameter sets, G3 and IOPB-I, known for their contrasting stiffness in describing the equation of state (EOS) of dense nuclear matter. The model incorporated three critical free parameters, namely, the transition density, which governs the onset of quark-nucleon phase, which defines the energy scale at which quark confinement occurs and the dark matter (DM) momentum, characterizing the properties of DM particles interacting with the quarkyonic matter. Each of these parameters plays a pivotal role in shaping the curvature distribution within the star, influencing both the central core and the outer layers. By systematically varying these parameters, we analyzed their effects on the EOS stiffness, mass-radius relation, and the radial profiles of curvature. The inclusion of dark matter introduced significant modifications to the curvature behaviour, redistributing mass-energy and altering the pressure and density gradients. This study highlights the interplay between nuclear physics, quark matter, and dark matter in determining the fundamental properties of compact stars, offering a more comprehensive understanding of their internal structure and the extreme conditions governing them.

Furthermore, studying the Kretschmann scalar (${\cal{K}}(r)$) and Weyl tensor ($\cal{W}$) curvature are significant as they only exist inside the neutron star. The Weyl tensor ($\cal{W}$) attains its maximum value at the surface as the density is in a diffused state. These analyses are more significant for the core region than the crust. Given that the crust's EOS remains poorly constrained, this region offers an ideal site for probing deviations from general relativity in the strong-field gravity regime, as well as studying phenomena like pulsar glitches. In the case of the Kretschmann scalar (${\cal{K}}(r)$), we observe an inverse relationship between the stiffness of the equation of state (EOS) and curvature patterns. The quarkyonic EOS is stiffer than baryonic and further softens after adding the dark matter. It highlights how the composition of matter affects the structure of neutron stars, with quark matter and dark matter playing key roles in shaping their compactness and curvature. Moreover, one can conclude that the quarkyonic stars are the most compact objects in nature and adding a certain amount of dark matter makes the star stable and meets the motivation. Notably, the maximum mass star dominates these effects more than the canonical mass star.

\section{Acknowledgments}
JAP acknowledges IOP for providing computer facilities and support from Science and Engineering Research Board (SERB) through the contingency of Ramanujan Fellowship File No. RJF/2022/000140.  
\bibliography{curvature}
\bibliographystyle{apsrev4-2}
\end{document}